\documentclass[10pt,twocolumn,aps,prd]{revtex4-1}
\usepackage{CJK}
\usepackage{amsmath,mathptmx,amssymb}
\usepackage{graphicx,graphics,subfigure}
\usepackage{ctable,multirow}
\usepackage{tabularx}
\usepackage[colorlinks=true,linkcolor=blue,citecolor=blue,urlcolor=blue]{hyperref}

\def\be{\begin{equation}}
\def\ee{\end{equation}}
\def\bea{\begin{eqnarray}}
\def\eea{\end{eqnarray}}
\def\ba{\begin{aligned}}
\def\ea{\end{aligned}}
\def\nn{\nonumber}
\def\p{\partial}

\begin{document}
\begin{CJK*}{GBK}{song}

\title{Topological classes of rotating black holes}

\author{Di Wu}
\email{wdcwnu@163.com}

\affiliation{School of Physics and Astronomy, China West Normal University,
Nanchong, Sichuan 637002, People's Republic of China}

\date{\today}

\begin{abstract}
In this paper, we investigate the topological numbers for the singly rotating Kerr black holes
in all dimensions and the four-dimensional Kerr-Newman black hole. We show that for uncharged
black holes, the rotation parameter has a significant effect on the topological number, and for
rotating black holes, the dimension of spacetimes has a remarkable effect on the topological
number too. In addition, we find that the topological numbers of the four-dimensional Kerr and
Kerr-Newman black holes are the same, which seems to indicate that the electric charge parameter
has no effect on the topological number of rotating black holes. Our current research provides
more evidence that the conjecture put forward in Wei \textit{et al.}
[\href{http://dx.doi.org/10.1103/PhysRevLett.129.191101}{Phys. Rev. Lett. \textbf{129}, 191101 (2022)}],
according to which all black hole solutions should be separated into three different topological
classes, is accurate, at least in the pure Einstein-Maxwell gravity theory.
\end{abstract}

\maketitle
\end{CJK*}

\section{Introduction}

As one of the most remarkable and fascinating objects in nature, the black hole has always been a
subject of extensive theoretical and observational studies. On the observational side, recent
years have witnessed remarkable successes in ``seeing" the shadow of a black hole via the event
horizon telescope (EHT) \cite{APJL875-L1,APJL930-L12} and ``hearing" gravitational waves from
black hole coalescence \cite{PRL116-061102,PRL116-241103}. On the theoretical side, the study
of the topology of black holes has very recently shed new light on the nature of gravity by
means of light rings \cite{PRL119-251102,PRL124-181101,PRD102-064039,PRD103-104031}, timelike
circular orbit \cite{2207.08397}, and thermodynamic properties \cite{PRL129-191101,PRD105-104003,
PRD105-104053,2207.02147,2207.10612,PRD106-064059,2208.10177,2211.05524,2211.12957,2211.15534}.

In particular, in Ref. \cite{PRL129-191101}, the black holes have been treated as topological
thermodynamic defects by using the generalized off-shell free energy, and thus black hole
solutions are divided into three different topological classes according to their different
topological numbers, which shed new light on the fundamental nature of quantum gravity. Because
of its simplicity and easy maneuverability of the procedure, the topological method proposed
in Ref. \cite{PRL129-191101} soon attracted a great deal of attention and was then successfully
applied to investigate the topological numbers of some known black hole solutions \cite{PRD106-064059,
2208.10177,2211.05524,2211.15534}, i.e., the Schwarzschild-AdS black hole solutions
\cite{PRD106-064059}, the static black hole solutions in Lovelock gravity \cite{2208.10177}, the static
Gauss-Bonnet-AdS black hole solutions \cite{2211.05524}, and the static black hole solution in
nonlinear electrodynamics \cite{2211.15534}. However, all of the above-mentioned progresses
\cite{PRL129-191101,PRD106-064059,2208.10177,2211.05524,2211.15534} are only restricted to the
static cases, leaving the topological numbers of the rotating black holes unexplored. Up to date,
the astronomical observation of black holes are basically rotating black holes, so it is very
important and necessary to study the topological number of rotating black holes, which also
provides the motivation of this paper.

In this paper, we investigate the topological numbers for the singly rotating Kerr black holes
in arbitrary dimensions and the four-dimensional Kerr-Newman black hole. We show that for
uncharged black holes, the rotation parameter has a significant effect on the topological
number, and for rotating black holes, the dimension of spacetimes has a remarkable effect
on the topological number too. The remaining part of this paper is organized as follows. In
Sec. \ref{II}, we first give a brief review of the topological approach, and then investigate
the topological number of the four-dimensional Kerr black hole. In Sec. \ref{III}, we extend
the discuss in Sec. \ref{II} to the cases of the $d$-dimensional singly rotating Kerr black
holes. In Sec. \ref{IV}, we turn to discuss the topological number of the four-dimensional
Kerr-Newman black hole. Finally, we present our conclusions in Sec. \ref{V}.

\section{Four-dimensional Kerr black hole}\label{II}

We first investigate the effect of the rotation parameter on the topological number of black
holes. We begin with by considering the four-dimensional Kerr black hole case \cite{PRL11-237},
whose metric in the Boyer-Lindquist coordinates has the form
\bea\label{Kerr}
ds^2 &=& -\frac{\Delta_r}{\Sigma}\big(dt -a\sin^2\theta\, d\phi \big)^2
 +\frac{\Sigma}{\Delta_r}dr^2 +\Sigma\, d\theta^2 \nn \\
&& +\frac{\sin^2\theta}{\Sigma}\big[adt -\big(r^2 +a^2\big)d\phi \big]^2 \,
\eea
where
\bea
\Delta_r = r^2 +a^2 -2mr \, , \quad \Sigma = r^2 +a^2\cos^2\theta \, , \nn
\eea
in which $m$ and $a$ are the mass and the rotation parameters of the black hole. The
thermodynamic quantities associated with the above solution (\ref{Kerr}) can be computed
via the standard method and have the following exquisite expressions:
\be\begin{aligned}\label{ThermKerr}
&M = m \, , \quad J = ma \, , \quad \Omega = \frac{a}{r_h^2 +a^2} \, , \\
&S = \pi(r_h^2 +a^2) \, , \quad  T = \frac{r_h^2 -a^2}{4\pi r_h(r_h^2 +a^2)} \, ,
\end{aligned}\ee
where $r_h = m \pm \sqrt{m^2 -a^2}$ are the locations of the event and Cauchy horizons.

With the expressions of the thermodynamical quantities in hand, we are now ready to investigate
the topological number of the four-dimensional Kerr black hole. As shown in Ref. \cite{
PRL129-191101}, one can first introduce the generalized off-shell free energy as
\be
\mathcal{F} = M -\frac{S}{\tau} \, ,
\ee
for a black hole system with mass $M$ and entropy $S$, where $\tau$ is an extra variable that
can be considered as the inverse temperature of the cavity surrounding the black hole. Only
when $\tau = 1/T$, the generalized free energy becomes on-shell.

In Ref. \cite{PRL129-191101}, a vector $\phi$ is established as
\be
\phi = \left(\frac{\p \mathcal{F}}{\p r_{h}}\, , \quad -\cot\Theta\csc\Theta\right) \, ,
\ee
where the parameter $0 \le \Theta \le \pi$ is introduced for convenience and intuition. The
component $\phi^\Theta$ is divergent at $\Theta = 0, \pi$, and the direction of the vector
points outward there. Furthermore, as discussed in Ref. \cite{2211.05524}, the extremal points
of the generalized free energy landscape exactly correspond to the on-shell black holes, so
the zero point of the component $\phi^{r_h}$ exactly meets the black hole solution. The
component $\phi^\Theta = 0$ will yield $\Theta = \pi/2$.

According to Duan's $\phi$-mapping topological current theory, a topological current can be
defined as \cite{SS9-1072,NPB514-705,PRD61-045004}
\be\label{jmu}
j^{\mu}=\frac{1}{2\pi}\epsilon^{\mu\nu\rho}\epsilon_{ab}\p_{\nu}n^{a}\p_{\rho}n^{b}\, , \quad
\mu,\nu,\rho = 0,1,2,
\ee
where $\p_{\nu} = \p/\p x^{\nu}$ and $x^{\nu}=(\tau,~r_h,~\Theta)$. The unit vector $n$ reads as
$n = (n^r, n^\phi)$, where $n^r = \phi^{r_h}/{||\phi||}$ and $n^\Theta = \phi^{\Theta}/{||\phi||}$.
It is a simple matter to prove that the above topological current (\ref{jmu}) is conserved, thus
one can easily derive $\p_{\mu}j^{\mu} = 0$. It is further shown that the topological current
$j^\mu$ is a $\delta$-function of the field configuration \cite{NPB514-705,PRD61-045004,
PRD102-064039}
\be
j^{\mu}=\delta^{2}(\phi)J^{\mu}\left(\frac{\phi}{x}\right)\, ,
\ee
where the three dimensional Jacobian $J^{\mu}\left(\phi/x\right)$ is defined as: $\epsilon^{ab}
J^{\mu}\left(\phi/x\right) = \epsilon^{\mu\nu\rho}\p_{\nu}\phi^a\p_{\rho}\phi^b$. It is easy to
see that $j^\mu$ equals to zero except when $\phi^a(x_i)$ = 0, and one can derive the expressions
of the topological number $W$ as:
\be
W = \int_{\Sigma}j^{0}d^2x = \sum_{i=1}^{N}\beta_{i}\eta_{i} = \sum_{i=1}^{N}w_{i}\, ,
\ee
where $\beta_i$ is the positive Hopf index counting the number of the loops of the vector $\phi^a$
in the $\phi$-space when $x^{\mu}$ are around the zero point $z_i$, $\eta_{i}=\textit{sign}(J^{0}
({\phi}/{x})_{z_i})=\pm 1$ is the Brouwer degree, and $w_{i}$ is the winding number for the $i$th
zero point of $\phi$ that is contained in $\Sigma$. It is worth to note that if two loops $\Sigma_1$
and $\Sigma_2$ surround the same zero point of $\phi$, then they have the same winding number. On
the other hand, if there is no zero point in the enclosed region, then one can have $W = 0$.

From the results already given above in Eq. (\ref{ThermKerr}), one can easily obtain the
generalized free energy
\be
\mathcal{F} = \frac{r_h^2 +a^2}{2r_h} -\frac{\pi(r_h^2 +a^2)}{\tau}
\ee
of the four-dimensional Kerr black hole. Then the components of the vector $\phi$ can be
computed as
\bea
&&\phi^{r_h} = \frac{1}{2} -\frac{a^2}{2r_h^2} -\frac{2\pi r_h}{\tau} \, , \\
&&\phi^{\Theta} = -\cot\Theta\csc\Theta \, .
\eea
By solving the equation $\phi^{r_h} = 0$, one can obtain its solution that is depicted by
a curve on the $r_h-\tau$ plane. For the four-dimensional Kerr black hole, one can get
\be
\tau = \frac{4\pi r_h^3}{r_h^2 -a^2} \, .
\ee

\begin{figure}[t]
\centering
\includegraphics[width=0.4\textwidth]{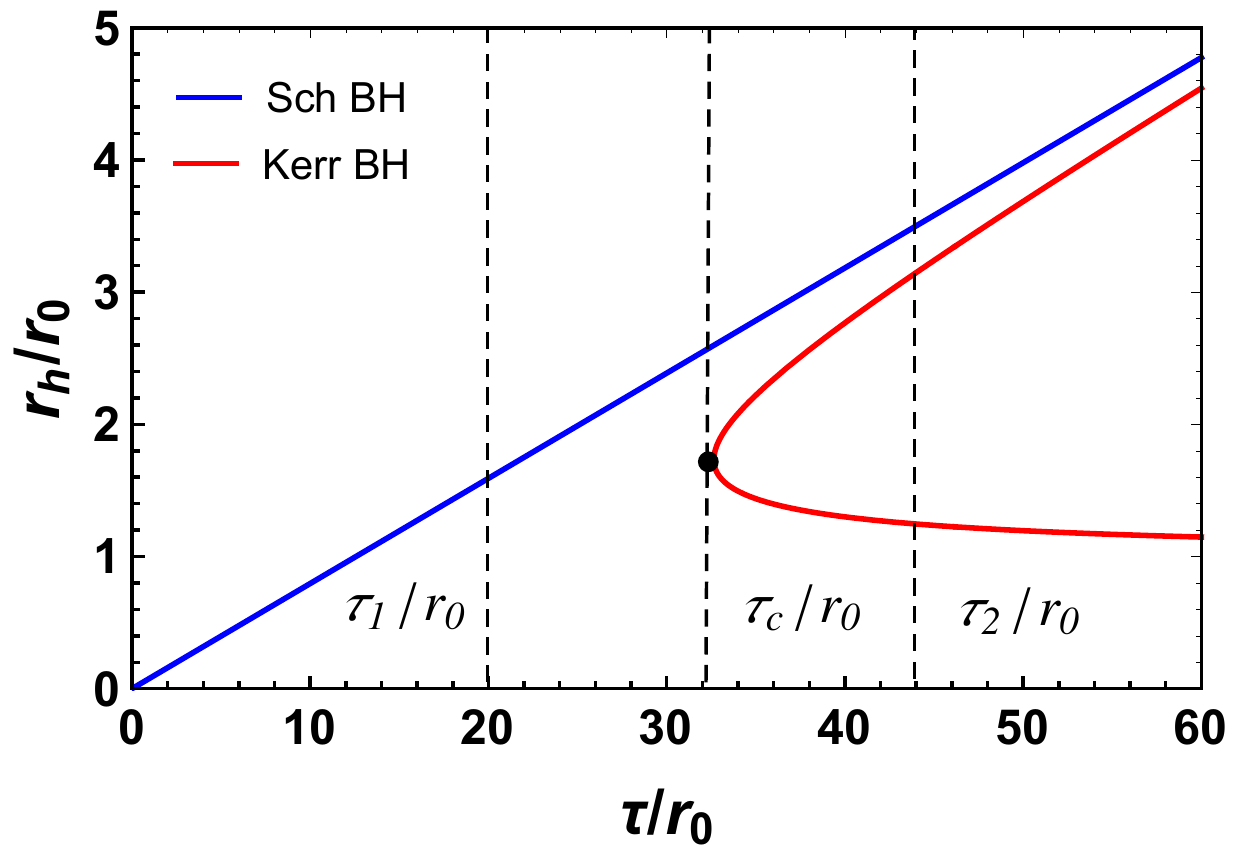}
\caption{Zero points of the vector $\phi^{r_h}$ are shown on the $r_h-\tau$ plane. To facilitate
comparison with previous results in Ref. \cite{PRL129-191101}, the blue and red solid lines
respond to the Schwarzschild black hole (Sch BH) and Kerr black hole (Kerr BH) with $a/r_0 = 1$,
respectively. The black dot with $\tau_c = 6\sqrt{3}\pi a$ represents the generation point for
the Kerr black hole. There is only one Schwarzschild black hole when $\tau = \tau_1$, while there
is one Schwarzschild black hole and two Kerr black holes when $\tau = \tau_2$.
\label{SchKerr}}
\end{figure}

\begin{figure}[t!]
\centering
\includegraphics[width=0.35\textwidth]{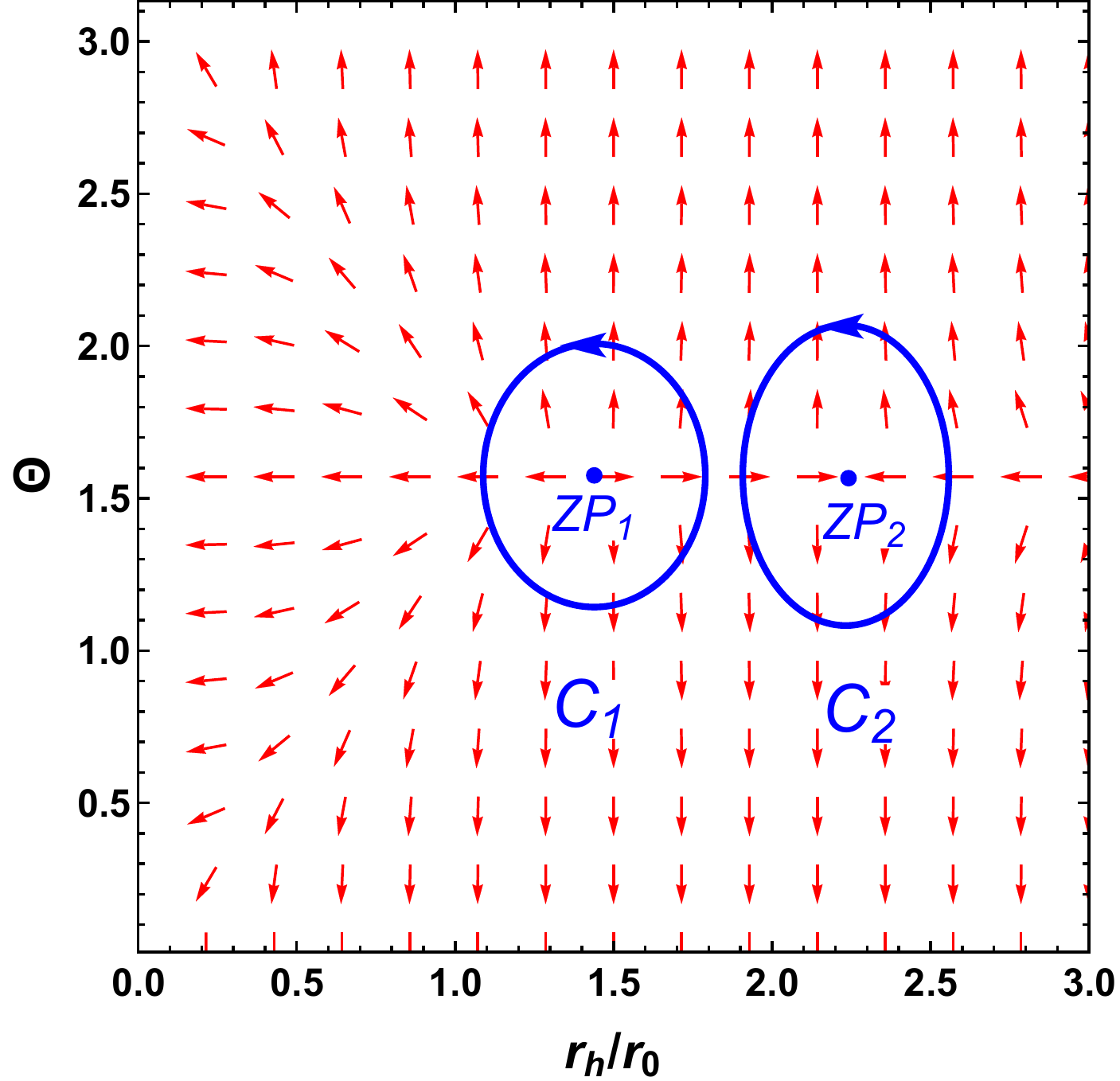}
\caption{The red arrows represent the unit vector field $n$ on a portion of the $r_h-\Theta$ plane.
The unit vector field is plotted for the four-dimensional Kerr black hole with $\tau/r_0 = 34.48$
and $a/r_0 = 1$. The zero points (ZPs) marked with blue dots are at $(r_h/r_0, \Theta) = (1.46,
\pi/2)$, and $(2.15, \pi/2)$, for ZP$_1$, and ZP$_2$, respectively. Due to the two winding numbers
$w_1 = 1$ and $w_2 = -1$, the topological number is: $W = w_1 +w_2 = 0$.
\label{4dKerr}}
\end{figure}

Fig. \ref{SchKerr} shows the zero points of the component $\phi^{r_h}$ for the four-dimensional
Kerr black hole with $a = r_0$, where $r_0$ is an arbitrary length scale set by the size of a
cavity enclosing the black hole. For large $\tau$, such as $\tau = \tau_2$, there are one and
two intersection points for the Schwarzschild and Kerr black holes, respectively. The intersection
points exactly satisfy the condition $\tau = 1/T$, and therefore represent the on-shell black
hole solutions with the characteristic temperature $T = 1/\tau$. Similar to the RN black hole
\cite{PRL129-191101}, but in contrast to the Schwarzschild black hole, the two intersection
points for the Kerr black hole can coincide each other when $\tau = \tau_c$, and then vanish
when $\tau < \tau_c$. Especially, at the point $\tau_c = 6\sqrt{3}\pi{} a$, one can easily
find $(d^2\tau/dr_h^2) = 6\sqrt{3}\pi/a > 0$ for the four-dimensional Kerr black hole. This
indicates that $\tau_c$ is a generation point, which can also be seen straightforward from
Fig. \ref{SchKerr}. Furthermore, the generation point $\tau_c$ divides the Kerr black hole
into the upper and lower branches with the winding number $w = -1$ and $w = 1$, respectively.
Thus one obtains the topological number $W = 0$ for the four-dimensional Kerr black hole.
Alternatively, one can also figure out the topological number for the four-dimensional Kerr
black hole by plotting the unit vector field $n$ for arbitrarily chosen typical values (Note
that $\tau$ must be larger than $\tau_c$), for example, $\tau/r_0 = 34.48$ and $a/r_0 = 1$
in Fig. \ref{4dKerr} where we find two zero points: ZP$_1$ at $r_h = 1.46r_0$ and ZP$_2$
at $r_h = 2.15r_0$, with the winding numbers $w_1 = 1$, $w_2 = -1$, respectively. So one can
get the topological number $W = w_1 +w_2 = 0$ for the four-dimensional Kerr black hole.
According to the classification proposal of black hole solutions, which is based upon their
different topological numbers \cite{PRL129-191101}, the Kerr black hole and RN black hole
are the same kind of black hole solutions since both of their topological numbers are equal
to zero. In addition, since the topological number of the Schwarzschild black hole is -1,
while that of Kerr black hole is 0, it implies that the rotation parameter has an important
effect on the topological number for the uncharged black hole.

\section{Higher dimensional singly rotating Kerr black holes}\label{III}

In this section, we will extend the above discussion to the cases of a rotating black hole in
higher dimensions by considering the $d$-dimensional singly rotating Kerr black holes. For the
singly rotating Kerr black holes in arbitrary dimensions, the metric has the form \cite{AP172-304,
PRD93-084015}
\bea\label{hdKerr}
ds^2 &=& -\frac{\Delta_r}{\Sigma}\left(dt -a\sin^2\theta d\phi \right)^2
 +\frac{\Sigma}{\Delta_r}dr^2 +\Sigma d\theta^2 \nn \\
&& +\frac{\sin^2\theta}{\Sigma}\left[adt -(r^2 +a^2)d\phi \right]^2
 +r^2\cos^2\theta d\Omega_{d-4}^2 \, , \qquad
\eea
where $d\Omega_{d}$ represents the line element of a $d$-dimensional unit sphere, and
\bea
\Delta_r = r^2 +a^2 -2mr^{5-d} \, , \quad \Sigma = r^2 +a^2\cos^2\theta \, . \nn
\eea
The thermodynamic quantities are \cite{PRD93-084015}
\be\ba\label{ThermhdKerr}
&M = \frac{d-2}{8\pi}\omega_{d-2}m \, , \quad J = \frac{\omega_{d-2}}{4\pi}ma \, ,  \\
&\Omega = \frac{a}{r_h^2 +a^2} \, , \quad
 S = \frac{\omega_{d-2}}{4}(r_h^2 +a^2)r_h^{d-4} \, , \\
&T = \frac{r_h}{2\pi}\left(\frac{1}{r_h^2 +a^2}
 +\frac{d-3}{2r_h^2} \right) -\frac{1}{2\pi r_h} \, ,
\ea\ee
where $\omega_{d-2} = 2\pi^{(d-1)/2}/\Gamma[(d-1)/2]$, and the horizon radius $r_h$ of the
black hole is determined by the equation: $\Delta_r = 0$.

From Eq. (\ref{ThermhdKerr}), one can get the generalized free energy
\bea
\mathcal{F} &=& M -\frac{S}{\tau} \nn \\
&=& \frac{\omega_{d-2}(d-2)(r_h^2 +a^2)}{16\pi r_h^{5-d}}
 -\frac{\omega_{d-2}(r_h^2 +a^2)r_h^{d-4}}{4\tau} \, . \quad
\eea
The components of the vector $\phi$ can be calculated as
\bea
\phi^{r_h} &=& \frac{\omega_{d-2}r_h^{d-6}}{16\pi\tau}\Big\{(d-2)\big[(d-3)\tau
 -4\pi r_h \big]r_h^2 \nn \\
&& +a^2\big[\tau(d-2)(d-5)-4(d-4)\pi r_h \big] \Big\}  \, , \\
\phi^{\Theta} &=& -\cot\Theta\csc\Theta \, .
\eea
By solving the equation $\phi^{r_h} = 0$, one can obtain
\be
\tau = \frac{4\pi[(d-2)r_h^3 +(d-4)a^2r_h]}{(d-2)[(d-3)r_h^2 +(d-5)a^2]}
\ee
as the zero point of the vector field $\phi$.

\begin{figure}[t]
\centering
\includegraphics[width=0.4\textwidth]{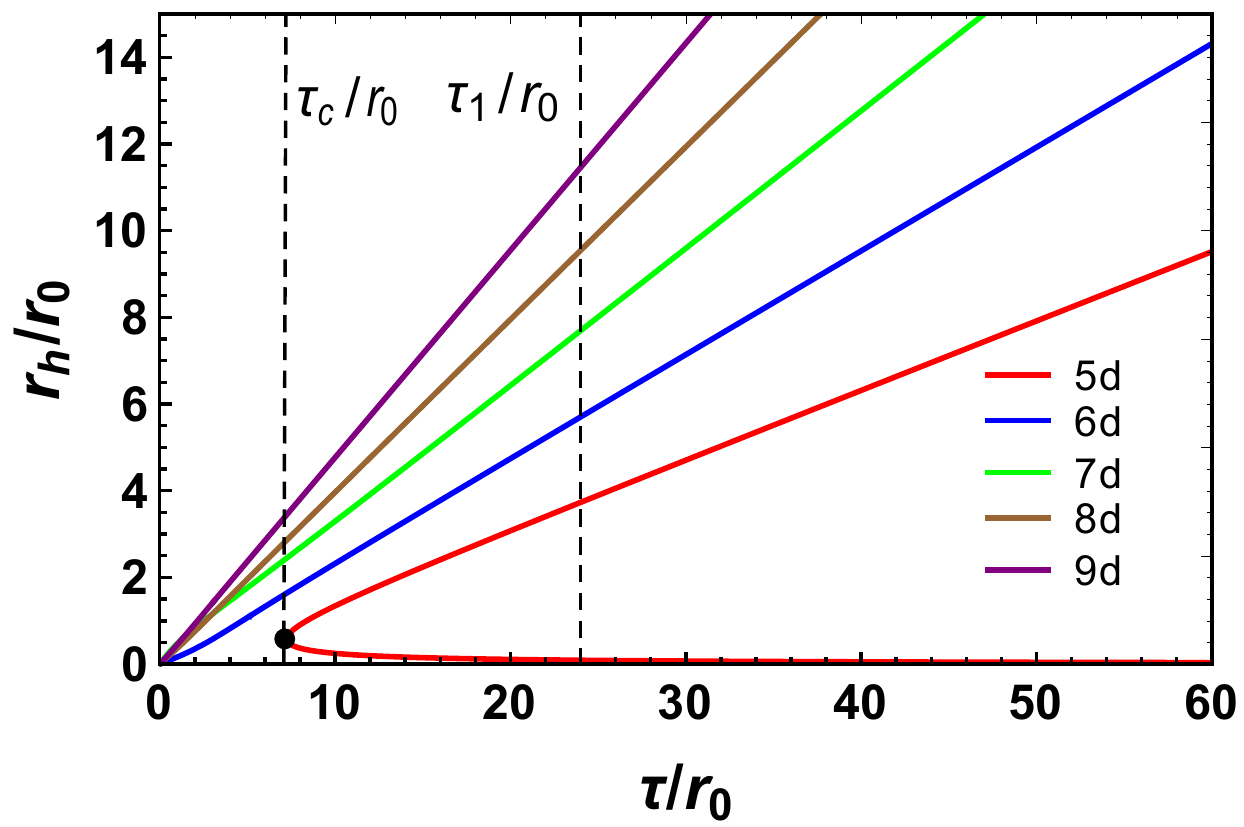}
\caption{Zero points of the vector $\phi^{r_h}$ are shown on the $r_h-\tau$ plane for five- to
nine-dimensional singly rotating Kerr black holes. The red, blue, green, brown, and purple solid
lines are for five- to nine-dimensional singly rotating Kerr black holes with $a/r_0 = 1$,
respectively. The black dot with $\tau_c = 4\sqrt{3}\pi a/3$ represents the generation point
for the five-dimensional singly rotating Kerr black hole.
\label{HdKerr}}
\end{figure}

\begin{figure}
\subfigure[~The unit vector field for the five-dimensional singly rotating Kerr black hole with
$\tau/r_0 = 20$ and $a/r_0 = 1$.]{\label{5dKerr}
\includegraphics[width=0.35\textwidth]{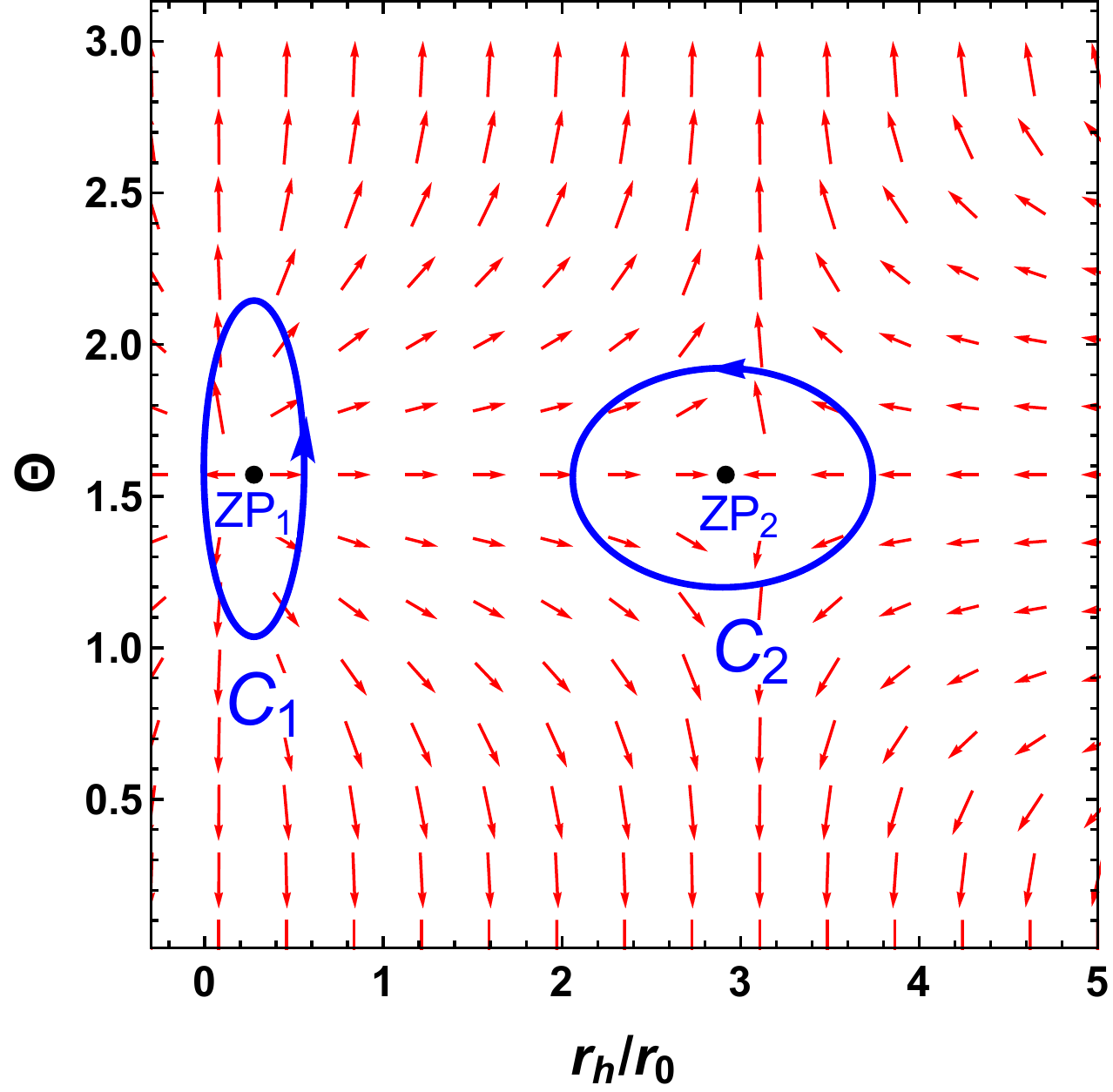}}
\subfigure[~The unit vector field for the six-dimensional singly rotating Kerr black hole with
$\tau/r_0 = 3\pi/2$ and $a/r_0 = 1$.]{\label{6dKerr}
\includegraphics[width=0.35\textwidth]{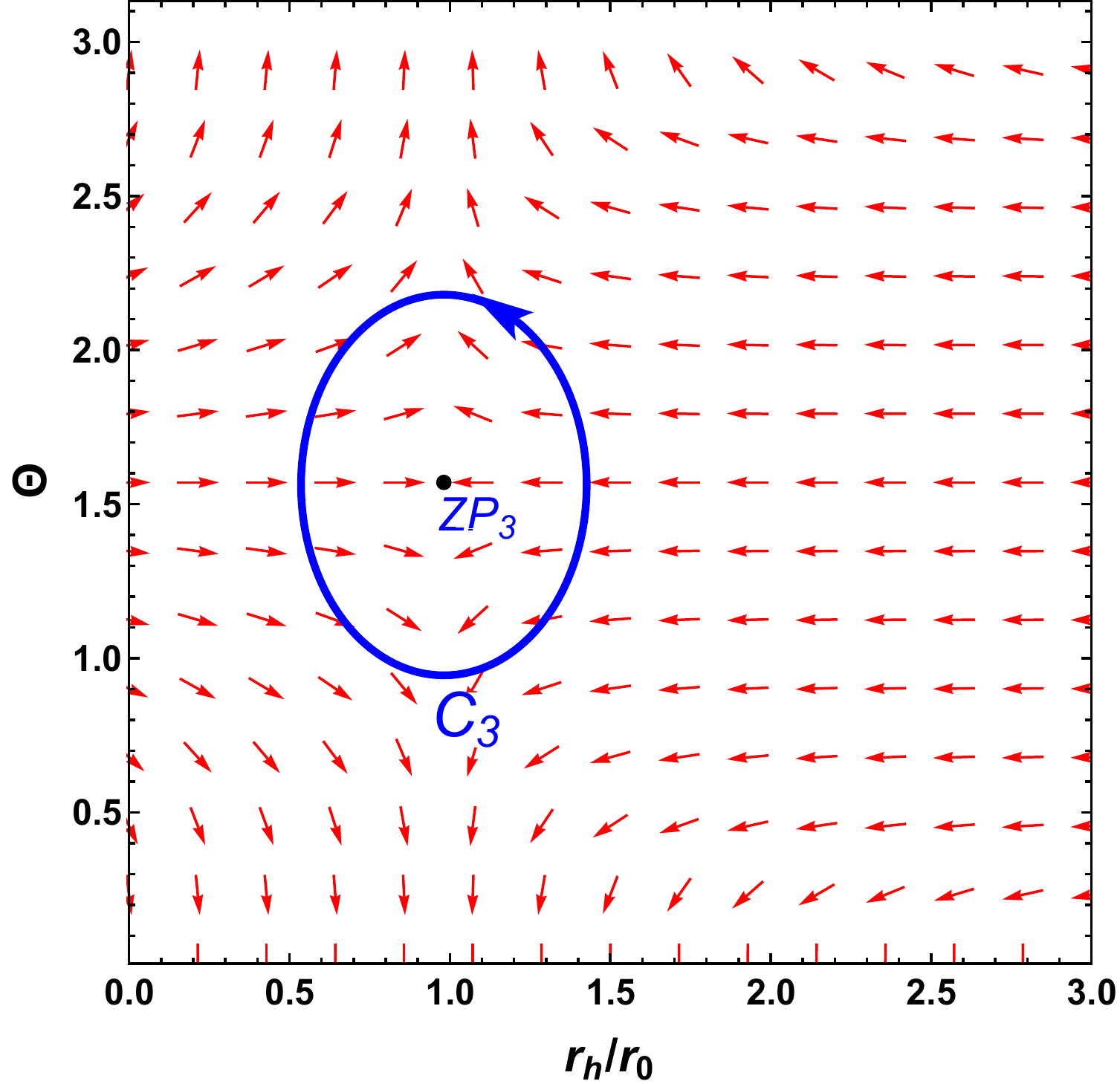}}
\subfigure[~The unit vector field for the seven-dimensional singly rotating Kerr black hole
with $\tau/r_0 = 16\pi/15$ and $a/r_0 = 1$.]{\label{7dKerr}
\includegraphics[width=0.35\textwidth]{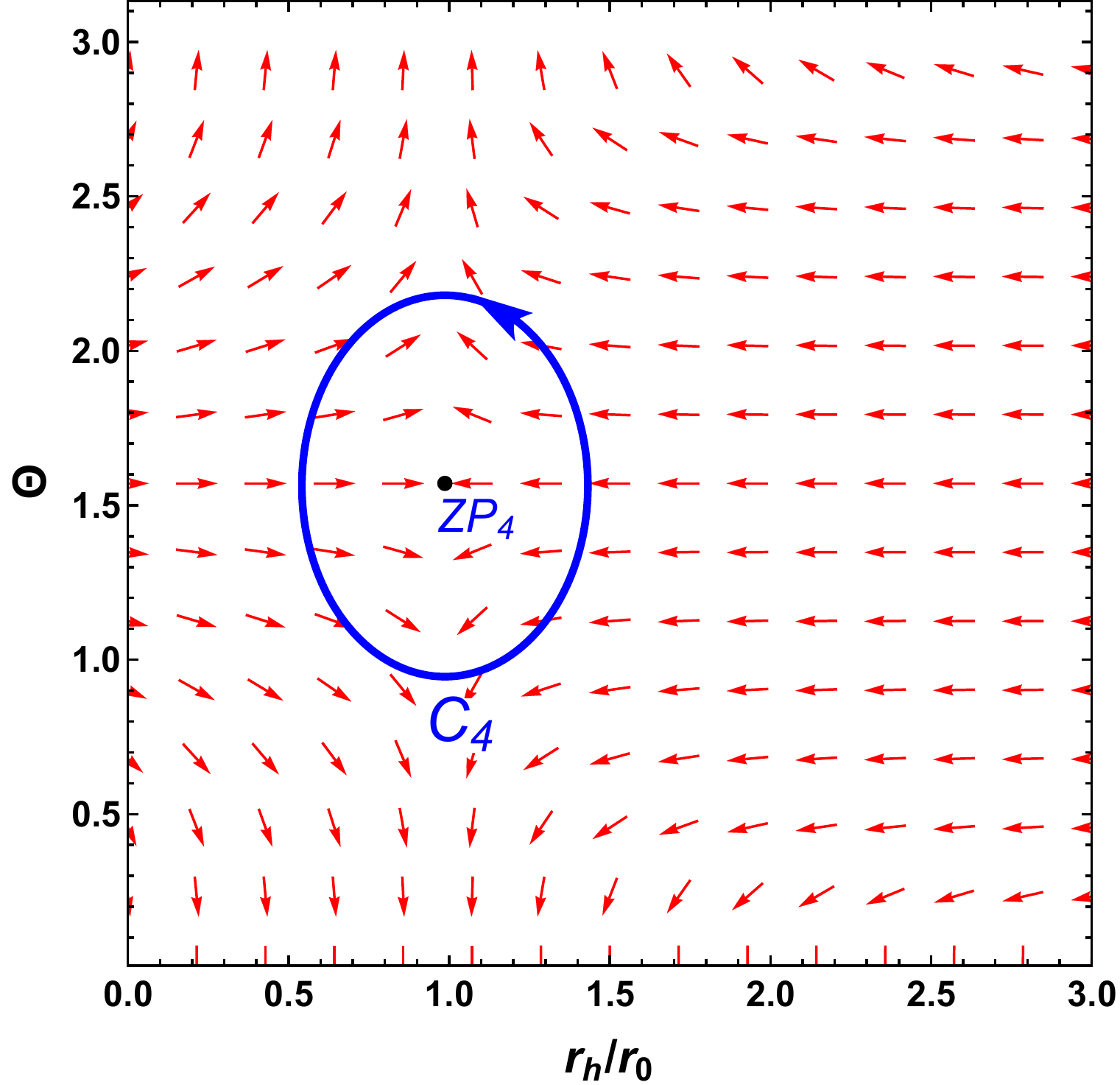}}
\caption{The red arrows represent the unit vector field $n$ on a portion of the $r_h-\Theta$
plane. The zero points (ZPs) marked with black dots are at $(r_h/r_0, \Theta) = (0.28,\pi/2)$,
$(3.14,\pi/2)$, $(1,\pi/2)$, and $(1,\pi/2)$, for ZP$_1$, ZP$_2$, ZP$_3$, and ZP$_4$, respectively.
The blue contours $C_i$ are closed loops surrounding the zero points.}\label{HdKerrvector}
\end{figure}

As some examples, we show the zero points of the component $\phi^{r_h}$ for five- to nine-dimensional
singly rotating Kerr black holes with $a = r_0$ in Fig. \ref{HdKerr}, and the unit vector field
$n$ for five- to seven-dimensional singly rotating Kerr black holes in Fig. \ref{HdKerrvector}
with $a = r_0$, $\tau = 20r_0 ~(d = 5)$, $3\pi{} r_0/2 ~(d = 6)$, and $16\pi{} r_0/15 ~(d = 7)$,
respectively. It is worthwhile to note that the unit vector field $n$ for eight- and nine-dimensional
singly rotating Kerr black holes is similar to that for six- and seven-dimensional singly rotating
Kerr black holes. The only difference is that the typical values are $\tau = 5\pi/6 ~(d = 8)$ and
$\tau = 24\pi/35 ~(d = 9)$, but here we are not going to plot them again.

From Figs. \ref{HdKerr} and \ref{HdKerrvector}, one can obtain the topological numbers of the higher
dimensional singly rotating Kerr black holes. For instance, the topological number of the
five-dimensional singly rotating Kerr black hole is $W = 1 -1 = 0$, which is the same one as that
of the four-dimensional Kerr black hole. However, via comparing Figs. \ref{SchKerr}-\ref{HdKerrvector},
the topological numbers of all $d \ge 6$ singly rotating Kerr black holes are found to be equal
to -1, which are different from the four- and five-dimensional Kerr black holes. To summarize,
according to the topological approach proposed in Ref. \cite{PRL129-191101}, we find that the
$d \ge 6$ singly rotating Kerr black holes and the $d = 4, 5$ singly rotating Kerr black holes
belongs to two different topological classes. Therefore, the dimension of spacetimes has an
important effect on the topological number of the rotating black holes, which are also reported
in Ref. \cite{2208.10177} for the higher-dimensional static uncharged black holes in Lovelock
gravity theory.

\section{Four-dimensional Kerr-Newman black hole}\label{IV}

Finally, we want to explore the effect of the electric charge parameter on the topological number
of the four-dimensional rotating black holes in the pure Einstein-Maxwell gravity theory. So in
this section, we turn to investigate the topological number of the four-dimensional Kerr-Newman
black hole \cite{JMP6-915,JMP6-918}, whose metric and Abelian gauge potential are
\bea
ds^2 &=& -\frac{\Delta_r}{\Sigma}\left(dt -a\sin^2\theta d\phi \right)^2
 +\frac{\Sigma}{\Delta_r}dr^2 +\Sigma d\theta^2 \nn \\
&&+\frac{\sin^2\theta}{\Sigma}\left[adt -(r^2 +a^2)d\phi \right]^2 \, , \\
A &=& \frac{qr}{\Sigma}(dt -a\sin^2\theta d\phi) \, ,
\eea
where
\bea
\Delta_r = r^2 +a^2 -2mr +q^2 \, , \quad \Sigma = r^2 +a^2\cos^2\theta \, . \nn
\eea
In the above, $m$ is the mass parameter, $a$ and $q$ are the rotation and electric charge
parameters, respectively.

The thermodynamic quantities can be evaluated via the standard method as follows:
\be\ba\label{ThermKN}
&M = m \, , \qquad J = ma \, , \qquad \Omega = \frac{a}{r_h^2 +a^2} \, , \\
&Q = q \, , \qquad \quad  \Phi = \frac{qr}{r_h^2 +a^2} \, , \\
&S = \pi(r_h^2 +a^2) \, , \qquad T = \frac{r_h^2 -a^2 -q^2}{4\pi r_h(r_h^2 +a^2)} \, ,
\ea\ee
where $r_h = m \pm \sqrt{m^2 -a^2 -q^2}$ are the locations of the outer and inner horizons.

Using the results given by Eq. (\ref{ThermKN}), one can directly obtain the generalized free
energy of the Kerr-Newman black hole as
\be
\mathcal{F} = M -\frac{S}{\tau} = \frac{r_h^2 +a^2 +Q^2}{2r_h} -\frac{\pi(r_h^2 +a^2)}{\tau} \, ,
\ee
and the components of the vector $\phi$ can be computed as
\bea
&&\phi^{r_h} = 1 -\frac{r_h^2 +a^2 +Q^2}{2r_h^2} -\frac{2\pi r_h}{\tau} \, , \\
&&\phi^{\Theta} = -\cot\Theta\csc\Theta \, .
\eea
One can also get
\be
\tau = \frac{4\pi r_h^3}{r_h^2 -a^2 -Q^2}
\ee
as the zero point of the vector field $\phi$.

\begin{figure}[t!]
\centering
\includegraphics[width=0.4\textwidth]{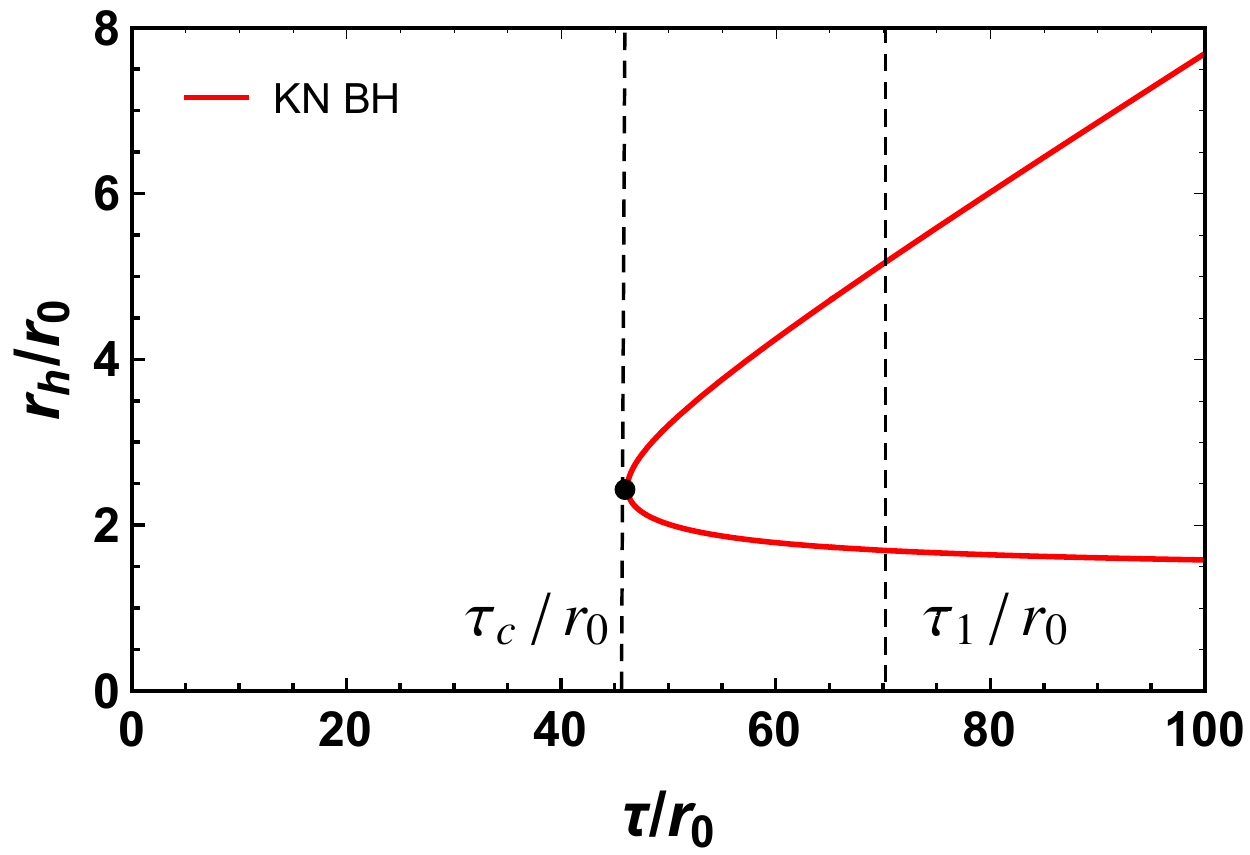}
\caption{Zero points of the vector $\phi^{r_h}$ are shown on the $r_h-\tau$ plane for the
Kerr-Newman black hole, with $a/r_0 = 1$ and $Q/r_0 = 1$. The black dot with $\tau_c = 6\pi
\sqrt{3(a^2 +Q^2)}$ denotes the generation point for the Kerr-Newman black hole. At $\tau =
\tau_c$, there is only one Kerr-Newman black hole, while at $\tau = \tau_1$, there are two
Kerr-Newman black holes.
\label{KerrNewman}}
\end{figure}

\begin{figure}[h]
\centering
\includegraphics[width=0.35\textwidth]{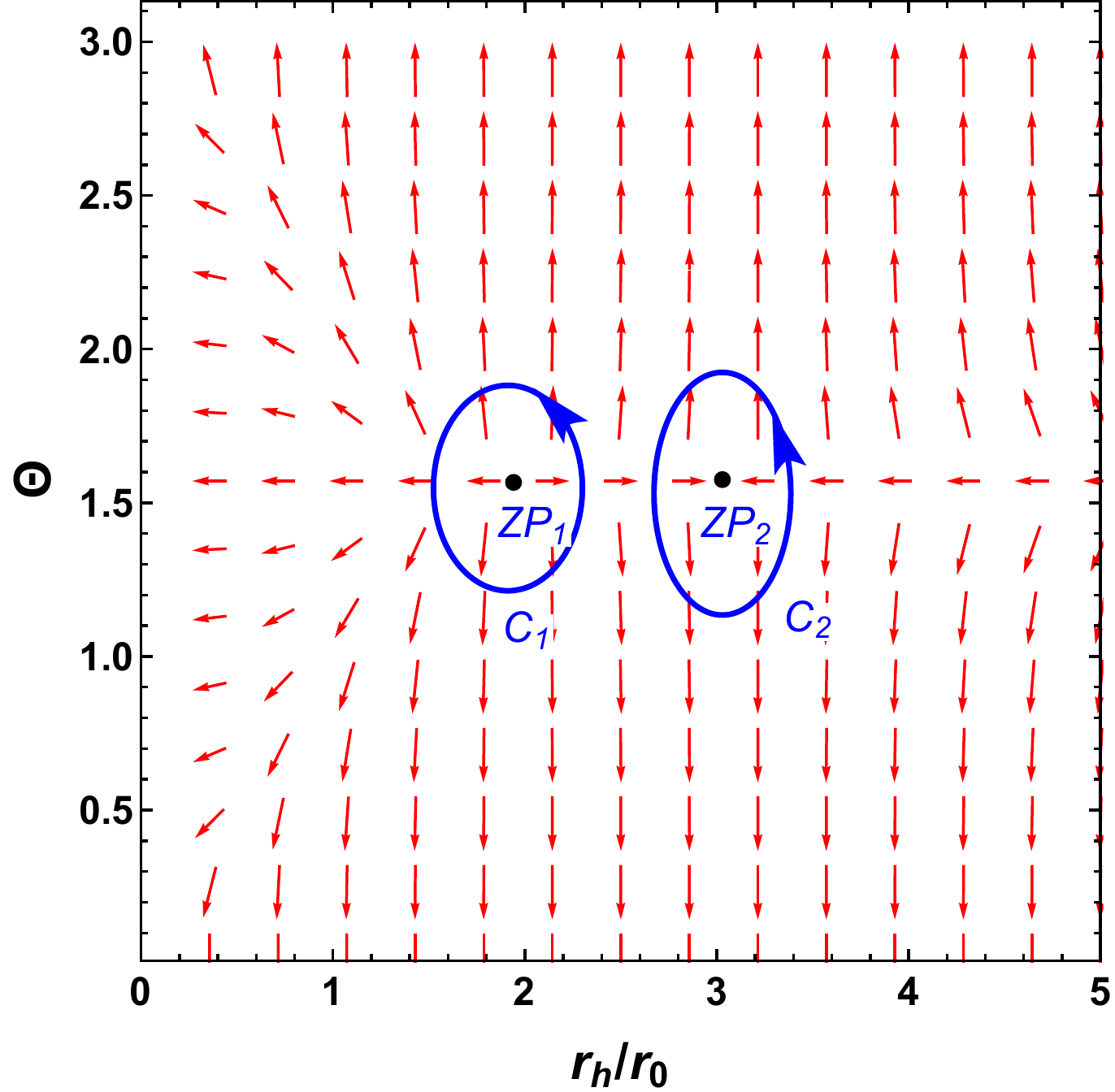}
\caption{The red arrows denote the unit vector field $n$ on a portion of the $r_h-\Theta$ plane.
The unit vector field is plotted for the Kerr-Newman black hole with $\tau/r_0 = 50$, $a/r_0 = 1$,
and $Q/r_0 = 1$. The zero points (ZPs) marked with black dots are at $(r_h/r_0, \Theta) = (1.95,
\pi/2)$, and $(3,\pi/2)$, for ZP$_1$, and ZP$_2$, respectively.
\label{4dKN}}
\end{figure}

Similar to the procedure adopted in the last two sections, we show the zero points of the component
$\phi^{r_h}$ with $a = r_0$ and $Q = r_0$ in Fig. \ref{KerrNewman}, and the unit vector field $n$
with $\tau = 50r_0$, $a = r_0$, and $Q = r_0$ in Fig. \ref{4dKN}. These two figures allow us to
determine that the topological number of the four-dimensional Kerr-Newman black hole is $W = 0$,
which is the same one as found for the four-dimensional Kerr black hole in Sec. \ref{II} and the
five-dimensional singly rotating Kerr black hole in Sec. \ref{III}. In addition, this also seems
to imply that the electric charge parameter has no effect on the topological number of rotating
black holes. However, this conclusion needs to be further tested by investigating the topological
numbers of many other rotating charged black holes.

\section{Conclusions}\label{V}

\begin{table}[h]
\caption{The topological number $W$, numbers of generation and annihilation points for various
black holes.}
\resizebox{0.48\textwidth}{!}{
\begin{tabular}{c|c|c|c}
\hline\hline
BH solutions &$W$ & Generation point & Annihilation point\\ \hline
Schwarzschild BH \cite{PRL129-191101} & -1 & 0 & 0\\
$d\ge 6$ singly rotating Kerr BH & -1 & 0 & 0\\ \hline
$d=5$ singly rotating Kerr BH & 0 & 1 & 0\\
$d=4$ Kerr BH & 0 & 1 & 0\\
Kerr-Newman BH & 0 & 1 & 0\\
RN BH \cite{PRL129-191101} & 0 & 1 & 0 \\ \hline
RN-AdS BH \cite{PRL129-191101} & 1 & 1 or 0 & 1 or 0\\
\hline\hline
\end{tabular}}
\label{I}
\end{table}

In this paper, we have investigated the topological numbers of the singly rotating Kerr black
holes in arbitrary dimensions and the four-dimensional Kerr-Newman black hole. Combined ours
with those in Ref. \cite{PRL129-191101}, Table \ref{I} summarizes some interesting results.
The $d \ge 6$ singly rotating Kerr black holes and the Schwarzschild black hole belong to
the first kind of topological classes since their topological number $W = -1$; the RN black
hole, the Kerr-Newman black hole, and the $d = 4,5$ singly rotating Kerr black holes belong
to the second kind of topological classes due to their topological number $W = 0$; the RN-AdS
black hole belong to the third kind of topological classes since its topological number
equals to 1.

In this work, we have arrived at two particularly exciting consequences: (i) the existence of the
rotation parameter has an important effect on the topological number of the uncharged black hole,
and (ii) the dimension of spacetimes has an important effect on the topological numbers of the
rotating black holes. In addition, the fact that the topological numbers of the four-dimensional
Kerr and Kerr-Newman black holes are identical seems to suggest that the electric charge parameter
has no effect on the topological number of the rotating black holes. Our current work also supports
the conjecture proposed in Ref. \cite{PRL129-191101} that all black hole solutions should be
classified into three different topological classes, at least in the pure Einstein-Maxwell
gravity theory.

There are two further promising topics to be pursued in the future. One intriguing topic is
to extend the present work to the more general cases with a nonzero cosmological constant, and
the black hole solutions in the supergravity and modified gravity theories. Another interesting
object is to investigate the topological number of the black hole solutions with the planar
\cite{PRD54-4891}, toroidal \cite{PRD56-3600}, and hyperbolic \cite{PRD92-044058}, as well as
spindle \cite{PRD89-084007,PRL115-031101,PRD102-044007,PRD103-044014,JHEP1121031} horizons to
explore whether there is a relation between the topological number and the horizon topology of
the black holes. We hope to report these related progress along these two directions soon.

\acknowledgments

We thank Prof. Shao-Wen Wei for very helpful discussions. This work is supported by the National
Natural Science Foundation of China (NSFC) under Grant No. 12205243, No. 11675130, by the Natural
Science Foundation of Sichuan Province under Grant No. 2023NSFSC1347, and by the
Doctoral Research Initiation Project of China West Normal University under Grant No. 21E028.


\begin{thebibliography}{99}
\def\JHEP{J. High Energy Phys.\,}
\def\PRD{Phys. Rev. D\,}
\def\PRL{Phys. Rev. Lett.\,}
\def\NPB{Nucl. Phys. B \,}
\def\PLB{Phys. Lett. B \,}
\def\JMP{J. Math. Phys. (N.Y.)\,}
\def\AP{Ann. Phys. (N.Y.)\,}
\def\APJL{Astrophys. J. Lett.\,}
\def\CPL{Chin. Phy. Lett.\,}

\bibitem{APJL875-L1}
The Event Horizon Telescope Collaboration, First M87 Event Horizon Telescope Results.
I. The Shadow of the Supermassive Black Hole,
\href{http://dx.doi.org/10.3847/2041-8213/ab0ec7}
{\APJL \textbf{875}, L1 (2019)}.

\bibitem{APJL930-L12}
The Event Horizon Telescope Collaboration, First Sagittarius A* Event Horizon Telescope Results.
I. The Shadow of the Supermassive Black Hole in the Center of the Milky Way,
\href{https://doi.org/10.3847/2041-8213/ac6674}
{\APJL \textbf{930}, L12 (2022)}.

\bibitem{PRL116-061102}
B.P. Abbott \emph{et al.} (LIGO Scientific and Virgo Collaborations),
Observation of Gravitational Waves from a Binary Black Hole Merger,
\href{http://dx.doi.org/10.1103/PhysRevLett.116.061102}
{\PRL \textbf{116}, 061102 (2016)}.

\bibitem{PRL116-241103}
B.P. Abbott \emph{et al.} (LIGO Scientific and Virgo Collaborations),
GW151226: Observation of Gravitational Waves from a 22-Solar-Mass Binary Black Hole Coalescence,
\href{http://dx.doi.org/10.1103/PhysRevLett.116.241103}
{\PRL \textbf{116}, 241103 (2016)}.

\bibitem{PRL119-251102}
P.V.P. Cunha, E. Berti, and C.A.R. Herdeiro,
Light Ring Stability in Ultra-Compact Objects,
\href{http://dx.doi.org/10.1103/PhysRevLett.119.251102}
{\PRL \textbf{119}, 251102 (2017)}.

\bibitem{PRL124-181101}
P.V.P. Cunha and C.A.R. Herdeiro,
Stationary Black Holes and Light Rings,
\href{http://dx.doi.org/10.1103/PhysRevLett.124.181101}
{\PRL \textbf{124}, 181101 (2020)}.

\bibitem{PRD102-064039}
S.-W. Wei,
Topological charge and black hole photon spheres
\href{https://doi.org/10.1103/PhysRevD.102.064039}
{\PRD \textbf{102}, 064039 (2020)}.

\bibitem{PRD103-104031}
M. Guo and S. Gao,
Universal properties of light rings for stationary axisymmetric spacetimes,
\href{https://doi.org/10.1103/PhysRevD.103.104031}
{\PRD \textbf{103}, 104031 (2021)}.

\bibitem{2207.08397}
S.-W. Wei and Y.-X. Liu, Topology of equatorial timelike circular orbits around stationary
black holes,
\href{https://arxiv.org/abs/2207.08397}{arXiv: 2207.08397}.

\bibitem{PRL129-191101}
S.-W. Wei, Y.-X. Liu, and R.B. Mann,
Black Hole Solutions as Topological Thermodynamic Defects,
\href{https://doi.org/10.1103/PhysRevLett.129.191101}
{\PRL \textbf{129}, 191101 (2022)}.

\bibitem{PRD105-104003}
S.-W. Wei and Y.-X. Liu,
Topology of black hole thermodynamics,
\href{https://doi.org/10.1103/PhysRevD.105.104003}
{\PRD \textbf{105}, 104003 (2022)}.

\bibitem{PRD105-104053}
P.K. Yerra and C. Bhamidipati,
Topology of black hole thermodynamics in Gauss-Bonnet gravity,
\href{https://doi.org/10.1103/PhysRevD.105.104053}
{\PRD \textbf{105}, 104053 (2022)}.

\bibitem{2207.02147}
M.B. Ahmed, D. Kubiznak, and R. B. Mann,
Vortex/anti-vortex pair creation in black hole thermodynamics,
\href{https://arxiv.org/abs/2207.02147}{arXiv: 2207.02147}.

\bibitem{2207.10612}
P.K. Yerra and C. Bhamidipati,
Topology of Born-Infeld AdS black holes in 4D novel Einstein-Gauss-Bonnet gravity,
\href{https://doi.org/10.1016/j.physletb.2022.137591}
{\PLB \textbf{835}, 137591 (2022)}.

\bibitem{PRD106-064059}
P.K. Yerra, C. Bhamidipati, and S. Mukherji,
Topology of critical points and Hawking-Page transition,
\href{https://doi.org/10.1103/PhysRevD.106.064059}
{\PRD \textbf{106}, 064059 (2022)}.

\bibitem{2208.10177}
N.-C. Bai, L. Li, and J. Tao,
Topology of black hole thermodynamics in Lovelock gravity,
\href{https://arxiv.org/abs/2208.10177}{arXiv: 2208.10177}.

\bibitem{2211.05524}
C.H. Liu and J. Wang,
The topological natures of the Gauss-Bonnet black hole in AdS space,
\href{https://arxiv.org/abs/2211.05524}{arXiv: 2211.05524}.

\bibitem{2211.12957}
Z.-Y. Fan,
Topological interpretation for phase transitions of black holes,
\href{https://arxiv.org/abs/2211.12957}{arXiv: 2211.12957}.

\bibitem{2211.15534}
C.X. Fang, J. Jiang, and M. Zhang,
Revisiting thermodynamic topologies of black holes,
\href{https://arxiv.org/abs/2211.15534}{arXiv: 2211.15534}.

\bibitem{PRL11-237}
R.P. Kerr,
Gravitational Field of a Spinning Mass as an Example of Algebraically Special Metrics,
\href{https://doi.org/10.1103/PhysRevLett.11.237}
{\PRL \textbf{11}, 237 (1963)}.

\bibitem{SS9-1072}
Y.-S. Duan and M.-L. Ge,
$SU$ (2) gauge theory and electrodynamics of $N$ moving magnetic monopoles,
\href{https://doi.org/10.1142/9789813237278_0001}
{Sci. Sin. \textbf{9}, 1072 (1979)}.

\bibitem{NPB514-705}
Y.-S. Duan, S. Li, and G.-H. Yang,
The bifurcation theory of the Gauss-Bonnet-Chern topological current and Morse function,
\href{https://doi.org/10.1016/S0550-3213(97)00777-3}
{Nucl. Phys. \textbf{B514}, 705 (1998)}.

\bibitem{PRD61-045004}
L.-B. Fu, Y.-S. Duan, and H. Zhang,
Evolution of the Chern-Simons vortices,
\href{https://doi.org/10.1103/PhysRevD.61.045004}
{\PRD \textbf{61}, 045004 (2000)}.

\bibitem{AP172-304}
R.C. Myers and M.J. Perry,
Black holes in higher dimensional space-times,
\href{https://doi.org/10.1016/0003-4916(86)90186-7}
{Ann. Phys. (N.Y.) \textbf{172}, 304 (1986)}.

\bibitem{PRD93-084015}
S.-W. Wei, P. Cheng, and Y.-X. Liu,
Analytical and exact critical phenomena of $d$-dimensional singly spinning Kerr-AdS black holes,
\href{https://doi.org/10.1103/PhysRevD.93.084015}
{\PRD \textbf{93}, 084015 (2016)}.

\bibitem{JMP6-915}
E.T. Newman and A.I. Janis,
Note on the Kerr spinning particle metric,
\href{https://doi.org/10.1063/1.1704350}
{J. Math. Phys. (N.Y.) \textbf{6}, 915 (1965)}.

\bibitem{JMP6-918}
E.T. Newman, E. Couch, K. Chinnapared, A. Exton, A. Prakash, and R. Torrence,
Metric of a rotating, charged mass,
\href{https://doi.org/10.1063/1.1704351}
{J. Math. Phys. (N.Y.) \textbf{6}, 918 (1965)}.

\bibitem{PRD54-4891}
R.G. Cai and Y.Z. Zhang,
Black plane solutions in four-dimensional space-times,
\href{http://dx.doi.org/10.1103/PhysRevD.54.4891}
{\PRD \textbf{54}, 4891 (1996)}.

\bibitem{PRD56-3600}
D.R. Brill and J. Louko,
Thermodynamics of (3+1)-dimensional black holes with toroidal or higher genus horizons,
\href{http://dx.doi.org/10.1103/PhysRevD.56.3600}
{\PRD \textbf{56}, 3600 (1997)}.

\bibitem{PRD92-044058}
Y. Chen, Y.K. Lim, and E. Teo,
Deformed hyperbolic black holes,
\href{http://dx.doi.org/10.1103/PhysRevD.92.044058}
{\PRD \textbf{92}, 044058 (2015)}.

\bibitem{PRD89-084007}
D. Klemm,
Four-dimensional black holes with unusual horizons,
\href{http://dx.doi.org/10.1103/PhysRevD.89.084007}
{\PRD \textbf{89}, 084007 (2014)}.

\bibitem{PRL115-031101}
R.A. Hennigar, R.B. Mann, and D. Kubiz\v{n}\'ak,
Entropy Inequality Violations from Ultraspinning Black Holes,
\href{http://dx.doi.org/10.1103/PhysRevLett.115.031101}
{\PRL \textbf{115}, 031101 (2015)}.

\bibitem{PRD102-044007}
D. Wu, P. Wu, H. Yu, and S.-Q. Wu,
Are ultraspinning Kerr-Sen-AdS$_4$ black holes always superentropic?,
\href{http://dx.doi.org/10.1103/PhysRevD.102.044007}
{\PRD \textbf{102}, 044007 (2020)}.

\bibitem{PRD103-044014}
D. Wu, S.-Q. Wu, P. Wu, and H. Yu,
Aspects of the dyonic Kerr-Sen-AdS$_4$ black hole and its ultraspinning version,
\href{http://dx.doi.org/10.1103/PhysRevD.103.044014}
{\PRD \textbf{103}, 044014 (2021)}.

\bibitem{JHEP1121031}
D. Wu and S.-Q. Wu,
Ultraspinning Chow's black holes in six-dimensional gauged supergravity and their
thermodynamical properties,
\href{http://dx.doi.org/10.1007/JHEP11(2021)031}
{\JHEP \textbf{11} (2021) 031}.

\end{thebibliography}
\end{document}